# STATISTICAL SURVEY AND ANALYSIS OF PHOTOMETRIC AND SPECTROSCOPIC DATA ON NEAS

**J. Rukmini[1], G. Raghavendra[2], S.A.Ahmed[1], D. Shanti Priya[1] and Syeda Azeem Unnisa[3]**
[1]*Department of Astronomy, Osmania University, Hyderabad*
[2]*Nizam College, Osmania University, Hyderabad*
[3]*Department of Environmental Science, Osmania University, Hyderabad*

**ABSTRACT**

Studies on Near Earth Asteroids (NEAs) throw light on discoveries, identification, orbit prediction and civil alert capabilities, including potential asteroid impact hazards. Due to various new observational programs, the discovery rate of NEAs has drastically increased over the last few years. In this paper we present the statistical survey and analysis of fundamental parameters (derived from Photometric and Spectroscopic observations) of a large sample of NEAs from various databases like IAU Minor Planet Center, European Asteroid Research Node (E.A.R.N.), Near Earth Objects - Dynamic Site (NEODyS-2), M4AST and SMASSMTT portals. We also discuss the characterization of NEAs on the basis of the correlations between the parameters studied from different observations and their physical implications in understanding the nature and physical properties of these objects.

## INTRODUCTION

Asteroids are rocky and metallic objects which are assumed to be left over pieces from the early solar system, 4.6 billion years ago. Asteroids have been discovered within the solar system and are hypothesized in the extra-solar planetary systems too. Johannes Kepler realized that the distance between the planets Mars and Jupiter was not proportional to the distances between other planets in the solar system. He concluded the presence of another undiscovered planet within this region (Kepler 1596). Around 20th century Otto Johannes Schmidt proposed that asteroids represented an arrested stage of planet formation which could not coalesce into a single large body. Today around 700,000 asteroids have been discovered with known orbits, giving insight into the Asteroid Belt's dynamic past. And measurements of the surface compositions have been studied for about 100,000 asteroids so far (Sanchez et al., 2001, Szabo et al., 2004, Nesvorny et al., 2005, Carvano et al., 2010).

Asteroids are often grouped on the basis of their orbital parameters. Most of the asteroids are located in the Main Belt, at heliocentric distances between 2.1 and 3.3 AU (these are called Main Belt Asteroids - MBAs). Some of the MBAs have

[1]corresponding author: rukminiouastro@yahoo.com

moved over time into the inner parts of the Solar System (Morbidelli *et al.*, 2002) after certain mechanisms like gravitational interactions with planets etc.

After the Tenguska event (1908), the Chelyabinsk event (Feb. 15, 2013) was recorded as the largest airburst of an asteroid in the earth's atmosphere causing havoc to the ecological balance on the planet (figures 1(a) & (b)). Majority of such objects, known as Near-Earth Objects (NEOs) reach the near-Sun state, strongly being affected by the heating and tidal processes at collisions and close approaches to Sun (as evident from shock blackened material and melt veins in the recovered fragments from the Chelyabinsk event (Vacheslav et al., 2014) seen in the inset of figure 1(b)). Small bodies of the Solar System are studied for scientific research, exploration of solar system objects, their origin and most importantly planetary protection

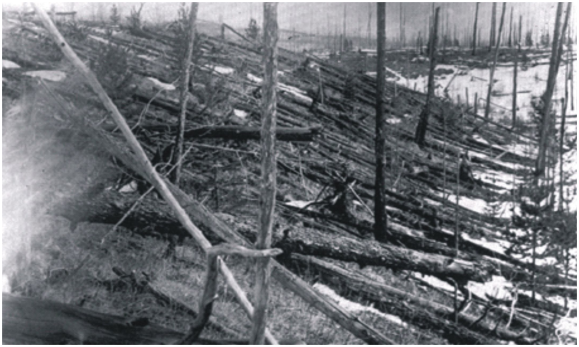

Fig 1(a)

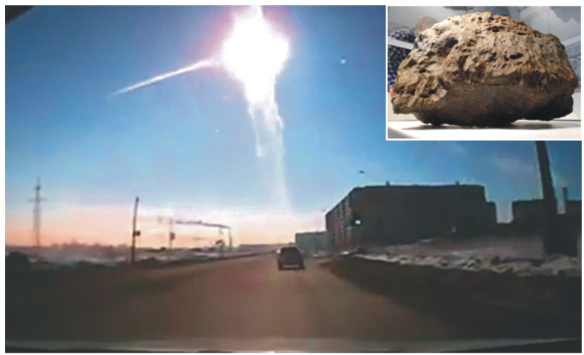

Fig 1(b)

**Figure 1(a):** The havoc raised by Tenguska event in 1908, that flattened the 2000 sq km of theforest in the explosion. Credit: the Leonid Kulik Expedition.
**Figure 1(b):** The Chelyabinsk explosion, the aftermath of which caused injuries to about 1,490 people.Credit: youtube.com

The immediate threat to Earth is defined by these objects, which are dynamically unstable and are continually sent into the inner solar system from reservoirs farther away from the Sun, containing rocky or icy bodies. Many ground and space based observations using telescope and radar are being used to discover, follow-up and characterize these potentially dangerous objects.

### Near-Earth Asteroids

(NEAs) are small bodies of the Solar System with perihelion distance '$q$' d" 1.3 AU (Astronomical Units) and aphelion distances '$Q$' e" 0.983 AU, whose orbits approach or intersect Earth orbit. Dynamical studies have confirmed the main belt origin for the majority of NEAs population. The transition of a main belt asteroid to NEA class is due to the dynamical perturbations associated with the main belt resonances created by the Jovian planets. Some of the most important issues in the study of NEOs are to understand their formation, possible relationship with meteorites and their role in the evolution of our solar system. NEOs have been broadly classified as minor planets/ asteroids (NEAs) and comets (NECs) on the basis of spectral/ orbital characteristics. The NEAs are classified into three main classes: Apollo, Amor and Aten on the basis of derived orbital parameters.

The three classes of NEAs are explained in the following.

### Atens

Semi-major axis '$a$' < 1.0 AU; '$Q$' e" 0.983AU – These are the Earth crossing asteroids with semi-major axes smaller than 1 AU. This means that they have orbital periods of less than a year and spend most of their time hidden by the Sun. They have sizes less than 3 km. Named after their prototype '2062 Aten', currently about 300 of them are known, some of which are also classified as Potentially Hazardous Asteroids (PHAs). Their expected lifetimes are relatively short, due to high probability of their collision with inner planets. This means that the population might not have formed in their current place but may be constantly replenished. And some of them could be definitely extinct.

*Apollos*

'a' e" 1.0 AU; 'q' d" 1.016 AU – These are the Earth crossing asteroids with semi-major axes larger than 1 AU. They have sizes less than 10 km. '1866 Sisyphus' is the largest discovered so far. They are named after their prototype '1862 Apollo'. About 1600 of them are known. They form the majority of the population of Earth- crossing asteroids and PHAs. Their expected lifetimes are also short, of the order of 10 million years due to their potential to collide with the inner planets, and hence they are also not formed in the current place but are constantly replenished. Although some are extinct cometary nuclei, it is currently understood that the majority have originated in the main asteroid belt and are ejected into the present orbits through gravitational interactions with Jupiter.

*Amors*

1.016 < 'q' < 1.3AU – These are Earth approaching asteroids with orbits that lie between the Earth and Mars, i.e. semi-major axes greater than 1 AU. While they do not cross the Earth's orbit, most of them do cross the orbit of Mars, and those with extreme eccentricities have their orbits beyond the orbit of Jupiter. There are currently around 1,500 of them catalogued, though close encounters with either Earth or Mars may alter their orbits, and make them evolve into Earth-crossing Apollo asteroids.

**PHAs** are currently defined based on parameters that measure the asteroid's potential to make threatening close approaches to the Earth. All asteroids with an Earth minimum orbit intersection distance (EMOID) smaller than 0.05 AU and an absolute magnitude (H) of 22.0 or brighter are considered PHAs (Milani et al., 2000). A sub-category of these asteroids are virtual impactors (VIs), objects for which the future Earth impact probability is non-zero according to the actual orbital uncertainty (Milani & Gronchi, 2010).

The NEA population shows a great diversity in terms of taxonomic classes, including almost all categories of asteroids, found in the Main Belt (Binzel et al, 2002). However the asteroid taxonomy was determined by a range of multi-band photometry and spectroscopy, such as the Small Main-belt Asteroid Spectroscopy Survey (SMASS). (e.g., Zellner et al. 1985; Xu et al. 1995; Bus & Binzel 2002a, 2002b; Lazzaro et al. 2004; de Leon et al. 2010). The detailed orbital parameters and spectroscopic properties of the different classes of NEOs are given in table 1 and table 2 respectively. Their study in addition holds photometric and spectroscopic clues on the possible mineralogical relationship between them and with that of other solar system objects.

*DATA COLLECTION*

The fundamental orbital parameters of a large sample of NEAs have been collected from various databases, like IAU Minor Planet Center, European Asteroid Research Node (E.A.R.N.) and Near Earth Objects - Dynamic Site (NEODyS-2), which include the data for 5522 Apollos, 4865 Amors, 964 Atens and 1636 PHAs. The fundamental parameters studied include absolute visual magnitude (H), eccentricity (e), semi-major axis (a), inclination (i), EMOID, perihelion (q) and aphelion distance (Q), albedo (p) and diameter (D), observed between 1932 – 2015.

The spectra of 197 Apollos, 160 Amors and 42 Atens, in Visual (V) and Infra-Red (IR) bands (0.45µm-2.45µm) used were obtained from observations done by Telescopio Nazionale Galileo of La Palam (TNG), European Southern Observatory, New Technology Telescope at chile, NASA Infrared Telescope Facility (IRTF) and also from campaigns like SMASS: Small Main-Belt Asteroid Spectroscopic Survey, MIT-UH-IRTF Joint Campaign for Spectral Reconnaissance which were made available through the different databases. Some of the data was collected from the observations done by DeMeo, F. E. et al. (2014), Thomson, C. A. et al. (2014), Binzel, R. P. et al. (2004), Somers, J. M. et al. (2010), Lazzarin, M. et al. (2004), Shepard, M. K. et al. (2008), Thomson, C. A. et al. (2011), Lazzarin, M. et al. (2010), Hicks, M. et al. (2012), Bus, S. J. & Binzel, R. P. (2002), Ieva, S. et al. (2014), Davies, J. K. et al. (2007), Whitely, R. J. (2001) and DeMeo, F. E. et al. (2009).

**Table 1:** Orbital parameters of NEAs

| ORBITAL PARAMETERS | APOLLO | AMOR | ATEN | PHAs |
|---|---|---|---|---|
| Perihelion distance (q) AU | q < 1.017 | 1.017 < q < 1.3 | 0.092 - 0.985* | 0.093 - 1.062* |
| Aphelion distance (Q) AU | 1.044 - 34.718* | 1.038 - 41.638* | Q > 0.983 | 0.955 – 35* |
| Semi-major axis(a)AU | a > 1.0 | a > 1.0 | a < 1.0 | 0.634 – 18* |
| Eccentricity (e) | 0.027 – 0.965 | 0.006 – 1* | 0.012 – 0.895 | 0.025 – 0.956 |
| Inclination (i)deg | 0 – 76* | 0.1 – 132* | 0 – 56.1* | - |
| Absolute Visual Magnitude (H) | 15 – 34* | 12 – 30* | 14.5 – 33* | H = 22.0 |
| EMOID (AU) | 0.00001 – 0.704* | 0.00038 – 0.705* | 0.0001 – 0.3901* | EMOID = 0.05 |
| Orbital period (yr) | 1 – 76* | 1.04 – 99.57* | 0.5 – 1* | 0.51 – 7.67* |
| Cross Earth Orbit or Not | Cross | Do Not Cross | Cross | Cross |
| **Albedo(p)** (using D = $1329*p^{-1/2}*10^{-0.2H}$) D – diameter, H – Absolute visual magnitude (Harris et al., 2002) | 0.09995 - .1003* | .09999 - .10001* | 0.09998 - .1001* | 0.09995 - .1003* |
| **Diameters (km)** | .0010 – 2.7767* | .0053 – 14.5722* | .0016 – 5.2908* | |
| **Orbital Image** | Earth / Apollo | Earth / Amor | Earth / Aten | |

*represents the values collected from the various databases for the current study.

**Table 2:** Spectroscopic classification and properties of NEAs.

| Taxonomy Type | Sub type/ Model Spectra | Definition Bus-DeMeo et al.(2009) | Relevant Composition from Burbine et al.(2002) |
|---|---|---|---|
| S -Complex | S, Sr, Sq, Sv spectra | -S (0.92 <Band min< 0.96) S or Sr ( Band min <0.92) - Sq (Band area < 0.016) | Minerals: Olivine, pyroxene. |
| C -Complex | C, Cb, Cg, Ch, B spectra | -B(-0.2<Slope< 0), -C (1 - 1.3 µm.) -Ch (0.7 µm), -Cg(0.55 µm), -Cgh(0.7 µm) | Minerals: Opaque's, Carbon, Phyllosilicates, some have weak features indicating Olivine, pyroxene. |
| X -Complex | X, Xe, Xc, Xk spectra | -X, Xc(Featureless) -Xe (Shows feature at 0.49µm), -Xk (feature between 0.8 - 1 µm ) | Minerals: M,P: Opaques, carbon, low-Fe pyroxene. E :enstatite, oldhamite. |
| End Members | L, Q, D, K, V spectra | -Q(Width≥0.25, Band area ≥0..16), | D:opaques,organics. K:Co,Cv,olivine. L : CAI- rich, spinel-rich O: Olivine, pyroxene. Q: Mostly LL OCs V:HEDs,pyroxene,plagioclase feldspar |

Table 3: Photometric (Peak Distribution) and Spectroscopic results for the sample of NEAs studied.

| Class of NEAs | Photometric Results (the range around peak distribution) | Spectroscopic Results |
|---|---|---|
| Apollo | Absolute Magnitude:20-22 & 24-26; Semi-major axis:1.0-1.4 AU; Eccentricity: 0.556-0.656; Inclination:0.1°-10°; Perihelion dist.:0.92-0.99 AU; Aphelion dist.:1.3-1.9 AU; EMOID:0.0001-0.05 AU; Diameter: 0.001-1.001 km; Albedo:0.09999-0.1 | S-complex found in majority followed by Q, C and X-complex; a very low number density noticed for other End members; Clear absence of M and F type observed |
| Amor | Absolute Magnitude:20-22 & 24-26; Semi-major axis:1.4-1.5 & 2.1-2.5 AU; Eccentricity: 0.406-0.606; Inclination:5°-10°; Perihelion dist.:1.0-1.1 AU; Aphelion dist.:3.0-3.4 AU; EMOID:0.05-0.1 AU; Diameter: 0.005-0.105 km; Albedo:0.099999-0.1 | S-complex found in majority followed by Q, C and X-complex; a very low number density in taxonomy class of V, K, D, M & F; Clear absence of O, L, EM type observed |
| Aten | Absolute Magnitude:20-22 & 24-26; Semi-major axis:1.4-1.5 & 2.1-2.5 AU; Eccentricity: 0.406-0.606; Inclination:5°-10°; Perihelion dist.:1.0-1.1 AU; Aphelion dist.:3.0-3.4 AU; EMOID:0.05-0.1 AU; Diameter: 0.005-0.105 km; Albedo:0.099999-0.1 | S-complex found in majority followed by Q, C and X-complex; a very low number density in taxonomy class of P, D, M, O & L; Clear absence of EM type observed |

**DISCUSSION**

The fundamental parameters were studied from statistical survey of photometric data to give frequency distributions of various orbital parameters. The spectra available in V and IR bands were visually studied and the taxonomic classes were determined. The conclusions derived from the above are listed for the three classes of NEAs in the table 3.

Additional Results: 1) e vs a : For the collected data there is clear distinction for Atens when compared to Amors and Apollos. Atens showed a decrease in 'e' with increase in 'a', while Amors and Apollos showed an increased distribution in 'e' with increase in 'a'. The cut-off in 'e' seems to be below 0.8 for Amors while, Apollos and Atens show the 'e' values till 1.0.

2) H vs EMOID: The distribution follows a similar trend for all the three classes with most of Atens showing minimum calculated EMOIDs; however all the three classes show a decreasing EMOIDs for increasing H (fainter objects).

Figures 2 & 3 show the characteristic dependence of e on a & H on EMOID respectively.

3) Figure 4 shows characteristic graph for different NEAs observed for the orbital parameters diameter (D taken in km) and frequency (f taken as 1/orbital period in hrs or the rotational rates). The rotation rates are determined mostly by the analysis of light curves (i.e. magnitude versus time plots). Since these parameters are observed over a large range, when studied for all the asteroids/minor planets including NEAs, they have been plotted on the logarithmic scales. The periods of more than 5500 minor planets are now known. The parameters of NEAs are over plotted on that of the known minor planets. The possible conclusions which were drawn from the plot are as follows.

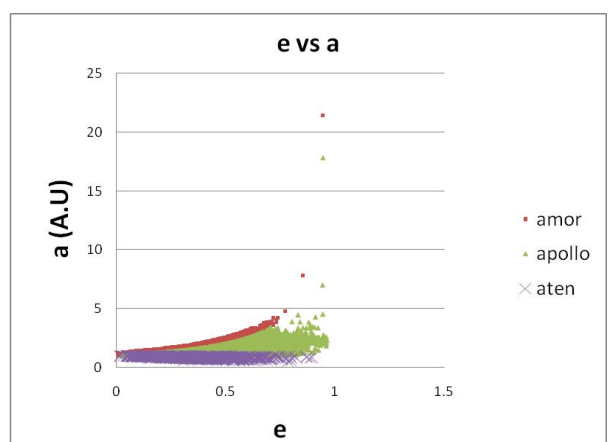

**Fig. 2:** The characteristic plot between eccentricity 'e' and semi-major axis 'a' drawn for the three classes of NEAs. The clear distinction of their orbital sizes seen for higher eccentricties.

Most of the minor planets including NEAs have rotation periods that are greater than 2.2 hours with the majority lying between 4 and 10 hours. For the current data on NEAs under study, the maximum period did not exceed 10 hours.

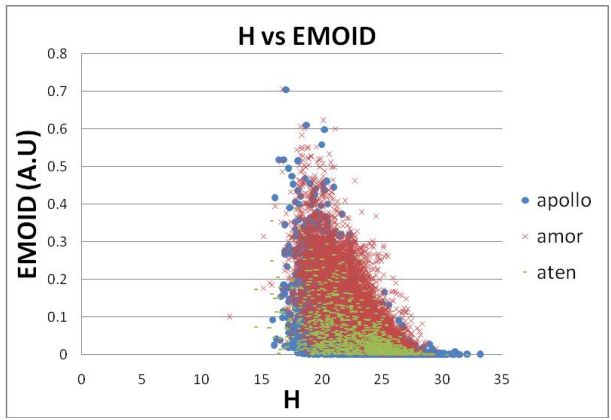

**Figure 3:** The characteristic plot between absolute visual magnitude 'H' and Earthminimum orbit intersection distance 'EMOID'.

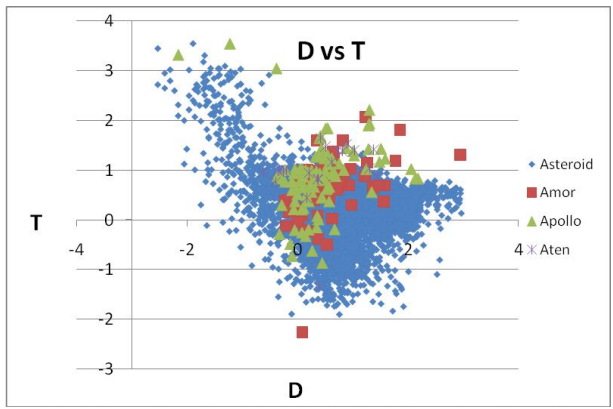

**Figure 4:** The characteristic plot between diameter 'D' and frequency of rotation 'T' taken in logarithmic scales

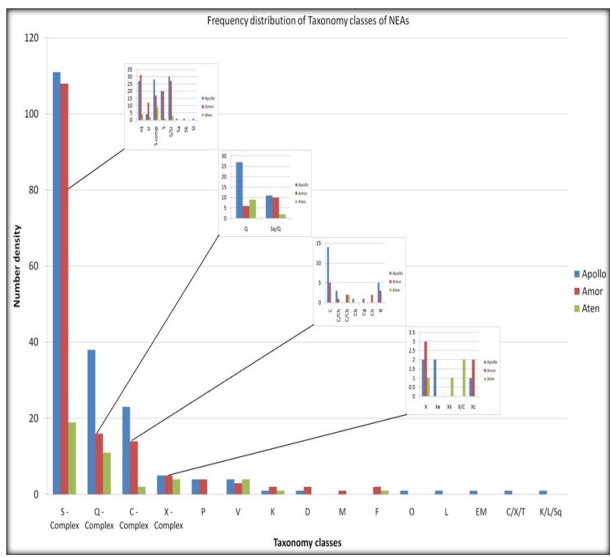

**Figure 5:** Frequency distribution of different taxonomy classes of NEAs with insets showing the distribution in each complex.

Asteroids spinning faster than 2.2 hours are "strength-bound" (like monolithic), it is interesting to note that some of the NEAs from all the three classes are found to exist in this region.

Smaller objects spinning slower than 2.2 hours are assumed to be "rubble piles", loose conglomeration of smaller pieces held together by mutual gravitation.

4) Spectroscopy: For the sample of data studied, S-complex asteroids are found to be the dominant class among all the three categories of NEAs, as seen in the frequency distribution of taxonomic classes in figure 5. This high proportion of S-complex could be in part due to selection effect or due to the fact that S-complex asteroids have higher albedos than C-complex making them easier to discover.

5) Most of the End members are only observed in Apollo ; M only observed in Amor; P are not seen in Aten objects-explained partly due to the source of their origin, which can be further substantiated by investigations of their mineralogical compositions.

Figure 5 shows the frequency distribution of various taxonomy complex types observed in the three categories of NEAs with the insets showing the distribution of sub-types in each taxonomy class.

## CONCLUSION

Billions of dollars are being spent in space research and for the search for extra-solar planets that are habitable. Failure to find one anywhere near vicinity indicates how precious our Earth, our home planet is and how careful we need to be to protect it and preserve it.

The discovery rate of NEOs has increased greatly over the last few years due to new observational programs. The study of the nature and physical properties of NEOs remains one of the priority researches in Solar system investigations which is necessary for addressing both fundamental scientific problems and its implications on habitability on this planet.

Recognizing the threat that NEOs pose to life on Earth, the US Congress has passed the 2005 NASA Authorization Act4 and NASA was directed to detect, track, catalogue, and characterize the physical characteristics of at least 90% of NEOs larger than 140 meters in diameter (H=22 mag) which can potentially impact the Earth by the end of year 2020 (Michael et al., 2015). Many amateur and professional groups are putting efforts to gather the fundamental parameters of these objects through ground and space based observations and building models to characterize them. Every analysis shall build up the preliminary database for these efforts and pave way for faster characterization of these objects.

Our knowledge of the structure and composition of NEAs is still limited, since only less than 10% of the known NEAs have orbital and spectral parameters determined from the observations. The current work adds statistical information to the ongoing efforts.

## ACKNOWLEDGEMENT

We thank the online resources provided by IAU Minor Planet Center, European Asteroid Research Node (E.A.R.N.) and Near Earth Objects - Dynamic Site (NEODyS-2) and SMASS II from where most of the data was collected and M4AST, which provided the tools for producing spectra for the collected data.